\documentclass{article}

\usepackage{authblk} 
\usepackage{ccaption} 
\usepackage{color}
\usepackage[T1]{fontenc} 
\usepackage[a4paper, total={6in, 8in}]{geometry}
\usepackage{graphicx}
\usepackage{hyperref}
\usepackage[normalem]{ulem}
\usepackage{rotating} 
\usepackage{multirow} 
\usepackage{array}
\usepackage{longtable} 
\usepackage{subfig}
\usepackage{xcolor}
\usepackage{marginnote}
\usepackage{soul}
\usepackage[textsize=scriptsize]{todonotes}

\usepackage[numbers]{natbib}

\makeatletter
\def\maxwidth{\ifdim\Gin@nat@width>\linewidth\linewidth\else\Gin@nat@width\fi}
\makeatother

\newcommand*{\email}[1]{\normalsize\href{mailto:#1}{#1}\par}

\author[1]{Christophe Vanderaa}
\author[1]{Laurent Gatto}
\affil[1]{Computational Biology and Bioinformatics Unit (CBIO), de
  Duve Institute, UCLouvain, Belgium}

\title{
    The current state of single-cell proteomics data analysis
}

\date{}

\begin{document}


\begin{centering}
    \maketitle
    Email: \email{laurent.gatto@uclouvain.be}
\end{centering}



\begin{abstract}


Sound data analysis is essential to retrieve meaningful biological
information from single-cell proteomics experiments. This analysis is
carried out by computational methods that are assembled into
workflows, and their implementations influence the conclusions that
can be drawn from the data. In this work, we explore and compare the
computational workflows that have been used over the last four years
and identify a profound lack of consensus on how to analyze
single-cell proteomics data. We highlight the need for benchmarking of
computational workflows, standardization of computational tools and
data, as well as carefully designed experiments. Finally, we cover the
current standardization efforts that aim to fill the gap and list the
remaining missing pieces, and conclude with lessons learned from the
replication of published single-cell proteomics analyses.

\end{abstract}






\section{Introduction}


Conducting a principled data analysis is not trivial, especially when
technologies and the data they generate increase in complexity at a
fast pace. This is particularly true for mass spectrometry (MS)-based
single-cell proteomics (SCP) data analysis. Several hurdles need to be
overcome in order to extract biologically meaningful information from
these complex data \citep{Vanderaa2021-ue}. Numerous methods exist to
correct for technical issues, and each method has its respective
advantages and drawbacks. In this review article, we show that the
variety of available methods to process proteomics data and the
current lack of computational standards has lead to a great
heterogeneity in SCP data analysis practices. This computational
heterogeneity is a reflection of the technical heterogeneity since
MS-based SCP has undergone many improvements. For instance, two sample
preparation strategies currently co-exist: SCP by label-free
quantification (LFQ) and multiplexed SCP \citep{Kelly2020-xd,
Slavov2021-hf, Ctortecka2021-bh}. Multiplexing strategies include
isobaric labelling, using tandem mass tags (TMT), or non-isobaric
labelling, using mass differential tags for relative and absolute
quantification (mTRAQ) \citep{Petelski2021-ah, Derks2022-yn}. Several
chips have been developed starting with the nanoPOTS chip
\citep{Zhu2018-bf}, followed by the N2chip \citep{Woo2021-th}, the
proteoCHIP \citep{Hartlmayr2021-ni}, or the microfluidic SciProChip
\citep{Gebreyesus2022-er}. Efforts have also focused on automation of
the sample processing and reported the successful integration of robot
handlers such as the Mantis \citep{Petelski2021-ah}, the OT-2
\citep{Liang2021-cr} or the CellenOne \citep{Hartlmayr2021-ni,
Leduc2022-cc, Woo2021-th} dispensing devices. Several MS instruments
have been used such as orbitraps or time of flight instruments
\citep{Orsburn2022-jt, Brunner2022-rd, Derks2022-yn} Furthermore, new
acquisition strategies are implemented such as data independent
acquisition mode \citep{Ctortecka2021-iq, Derks2022-yn,
Ctortecka2022-wm, Brunner2022-rd, Gebreyesus2022-er}, prioritized data
acquisition \citep{Gray_Huffman2022-cj}, or increased precursor
sampling and identification transfer \citep{Webber2022-vk, Woo2022-rt}
that all allow for reduced missing values. Finally, several groups
reported the acquisition of post-translational modifications, further
increasing the biological resolution of the technology
\citep{Orsburn2022-jt, Li2021-qv}. This technical heterogeneity is
thoroughly justified and benchmarked; each publication demonstrates
the added value of its experimental workflow. As the field
demonstrates its potential, efforts are made to make the technology
broadly accessible and standardized through detailed protocols
\citep{Petelski2021-ah, Leduc2022-cc, Gray_Huffman2022-cj,
Derks2022-yn} or by replacing custom-built material with commercially
available devices \citep{Liang2021-cr, Tsai2021-so}. Several groups
performed a thorough fine-tuning of experimental and instrumental
parameters to better understand their impact on analytical performance
\citep{Tsai2020-gp, Cheung2020-wc, Specht2021-ui}. The current state
of the field and the opportunities to push the SCP technology to its
full potential are regularly being discussed, sparking the interest of
a growing community \citep{Levy2018-bl, Specht2018-vs, Slavov2020-go,
Kelly2020-xd, Slavov2021-hf, Ctortecka2021-bh, Slavov2021-px,
Slavov2021-kk, Slavov2021-ab}. These efforts however mostly focus on
the technical aspects of the technology and overlook the current
computational practices.

In this review, we provide a computational perspective to the
discussion and examine the current approaches and practices for
analysing SCP data, specifically focusing on quantitative data
processing. The first section highlights the current heterogeneity in
SCP data processing. The next section covers the existing tools that
bring a solution to the current hurdles. Finally, the last section
provides several guidelines on how to improve SCP data analysis
practices.

\section{Quantitative data processing lacks consensus}


Proteomics data analysis encompasses three main tasks: spectral data
processing, quantitative data processing and downstream data analysis.
Spectral data processing identifies and quantifies the peptides from
the acquired MS spectra. Assigning peptide sequences to MS spectrum
was spotlighted as an important challenge for SCP data analysis
\citep{Slavov2021-ab} and several groups have contributed to
methodological and software improvements. For instance, Yu et al.\
extended the match between run (MBR) algorithm from MaxQuant to TMT
data, taking advantage of the quantification data present in
unidentified MS2 spectra \citep{Yu2020-wl}.  The \texttt{iceR} package
also propagates information across runs. The algorithm dramatically
improves peptide identification and outperforms MBR
\citep{Kalxdorf2021-sh}. Unfortunately, \texttt{iceR} is only
applicable to label-free data. Another approach to improve peptide
identification is to increase the confidence of matching by
re-scoring. Re-scoring uses the annotations generated by the search
engines such as the deviation between expected and measured elution
times or m/z, the peptide length, or the ion charge
\citep{Van_Der_Watt2021-mf}, to update the score or probability that
measured spectra correctly match spectra from a theoretical or
empirical spectral library. DART-ID, a Bayesian framework to update
posterior error probabilities based on an accurate estimation of
elution times, has been applied to SCP data and showed a significant
increase in the number of identified spectra \citep{Chen2019-uc}.
Others have also improved the Percolator re-scoring algorithm for SCP
experiments \citep{Fondrie2021-dc, Fondrie2020-yx}, although the
measured improvements were subtle. While these developments
considerably improve the quality of spectrum identification, no
dedicated developments in quantitative data processing have been
reported.


Quantitative data processing plays a critical role to overcome many
technical artefacts and to satisfy downstream analysis requirements.
It consists of several steps. Quality controls ensure the analysed
data are composed of reliable information and remove features of low
quality that could otherwise compromise the validity of the
results. Aggregation combines peptide level data into protein level
data. Log-transformation shapes the data so that the quantitative
values follow normal distributions. Imputation generates estimates for
missing values.  Finally, normalization and batch correction aim to
remove technical differences between samples and are essential to
avoid biased results.  Each of these steps is implemented using
different methods. For instance, many methods exist for missing value
imputation: replace by zero, replace with random values sampled from
an estimated background distribution, replace by values estimated from
the K-nearest neighbours (KNN),\ldots The imputation methods have
different underlying assumptions that have been extensively reviewed
in the bulk proteomics field \citep{Bramer2021-ml}, but further
research is required to assess whether these assumptions remain valid
or not for SCP data. Besides choosing the right method, finding a
correct sequence of steps is another challenge. For instance, batch
effects influence missing data and vice versa
\citep{Vanderaa2021-ue}. It has been suggested to correct for batch
effects before imputation \citep{Cuklina2021-kf}, but batch correction
methods such as ComBat \citep{Johnson2007-nc} break with highly
missing data as in SCP data.


As of today, developing computational workflows for SCP quantitative
data processing requires expert knowledge. We refer to ``computational
workflow'' or ``computational pipeline'' as the sequence of steps and
methods that process quantification data for downstream statistical
testing or visualization. Computational workflows are built from
scratch and their development often lacks an explicit rationale. Since
we lack systematic comparisons, benchmarks or guidelines, the
processing approaches become fundamentally different between
publications. To illustrate our claim, we review the computational
approaches from several studies that shaped the SCP landscape since
2018 (Table~\ref{experiment_table}). These studies present significant
contributions to the field and showcase applications on actual single
cells (as opposed to bulk lysate dilutions). Five studies supplemented
their publication with material allowing to repeat, at least
partially, their computational analysis. Three studies from the Slavov
Lab provide the R code and the data required to fully repeat their
results \citep{Specht2021-jm, Leduc2022-cc, Derks2022-yn}. The code is
however poorly documented and difficult to re-use by other labs.
Schoof et al.\ also offer the data used to repeat their study and
distribute their computational workflow as a documented python
library, \texttt{sceptre}~\citep{Schoof2021-pv}. Their library heavily
relies on \texttt{scanpy}, a popular python library for scRNA-Seq
analysis \citep{Wolf2018-xo}. Finally, Brunner et al.\ provide a
python script that also relies on \texttt{scanpy}, but it lacks an
explicit link with the input data \citep{Brunner2022-rd}. Based on the
available material (scripts for \citep{Specht2021-jm, Schoof2021-pv,
Brunner2022-rd, Leduc2022-cc, Derks2022-yn} or the methods section for
the others), we constructed Figure~\ref{fig:workflows_overview}. We
divide the workflow steps in 7 general categories and further group
the different steps depending on whether they are applied at the
precursor/peptide to spectrum match (PSM) level,
peptide level, protein level or implicitly embedded in an MS data
preprocessing software.

\begin{sidewaystable}
  \caption{ \textbf{Overview of influential SCP studies.} These
    studies were published between 2018 and 2022. MaxQuant, FragPipe,
    Proteome Discoverer (PD), and DIA-NN are software tools to conduct
    peptide identification and quantification. The peptide
    identification is performed by underlying search engines such as
    Andromeda, MS-GF+, MSFragger or SEQUEST.\ Multiplexing relies on
    TMT or mTRAQ labelling while no labelling implies an LFQ approach.
    Some publication link to associated computational scripts to
    reproduce the analysis that were written either in python or R.
    The throughput is expressed in number of cells retained after
    sample quality control, if any
    (Figure~\ref{fig:workflows_overview}A).\label{experiment_table}}
  \small
  \begin{tabular}{@{\extracolsep{5pt}} lllllll}
    \\[-1.8ex]
    \hline
    \hline
    \\[-1.8ex]
    Study & Publication date & Raw data analysis & Labeling & Script & Throughput & Reference \\
    \hline \\[-1.8ex]
    Zhu et al.\ 2018             &   Sep 2018    &   MaxQuant/Andromeda  &   ---             &   ---     &   6       &   \citep{Zhu2018-mi}      \\
    Budnik et al.\ 2018          &   Oct 2018    &   MaxQuant/Andromeda  &   TMT-10          &   ---     &   190     &   \citep{Budnik2018-qh}   \\
    Dou et al.\ 2019             &   Oct 2019    &   MS-GF+,MASIC        &   TMT-10          &   ---     &   72      &   \citep{Dou2019-wm}      \\
    Zhu et al.\ 2019             &   Nov 2019    &   MaxQuant/Andromeda  &   ---             &   ---     &   28      &   \citep{Zhu2019-ja}      \\
    Cong et al.\ 2020            &   Jan 2020    &   MaxQuant/Andromeda  &   ---             &   ---     &   4       &   \citep{Cong2020-zb}     \\
    Tsai et al.\ 2020            &   May 2020    &   MaxQuant/Andromeda  &   TMT-11          &   ---     &   104     &   \citep{Tsai2020-gp}     \\
    Williams et al.\ 2020, LFQ   &   Aug 2020    &   MaxQuant/Andromeda  &   ---             &   ---     &   17      &   \citep{Williams2020-ty} \\
    Williams et al.\ 2020, TMT   &   Aug 2020    &   MaxQuant/Andromeda  &   TMT-11          &   ---     &   152     &   \citep{Williams2020-ty} \\
    Liang et al.\ 2020           &   Dec 2020    &   FragPipe/MSFragger  &   ---             &   ---     &   3       &   \citep{Liang2021-cr}    \\
    Specht et al.\ 2021          &   Jan 2021    &   MaxQuant/Andromeda  &   TMT-11,TMT-16   &   R       &   1,490   &   \citep{Specht2021-jm}   \\
    Cong et al.\ 2021            &   Feb 2021    &   PD/SEQUEST          &   ---             &   ---     &   6       &   \citep{Cong2021-qa}     \\
    Schoof et al.\ 2021          &   Jun 2021    &   PD/SEQUEST          &   TMT-16          &   Pyhon   &   2,025   &   \citep{Schoof2021-pv}   \\
    Woo et al.\ 2021             &   Oct 2021    &   MaxQuant/Andromeda  &   TMT-16          &   ---     &   108     &   \citep{Woo2021-th}      \\
    Brunner et al.\ 2022         &   Feb 2022    &   DIA-NN              &   ---             &   Python  &   231     &   \citep{Brunner2022-rd}  \\
    Leduc et al.\ 2022           &   Mar 2022    &   MaxQuant/Andromeda  &   TMT-18          &   R       &   1,556   &   \citep{Leduc2022-cc}    \\
    Woo et al.\ 2022             &   Mar 2022    &   MaxQuant/Andromeda  &   ---             &   ---     &   155     &   \citep{Woo2022-rt}      \\
    Webber et al.\ 2022          &   Apr 2022    &   PD/SEQUEST          &   ---             &   ---     &   28      &   \citep{Webber2022-vk}   \\
    Derks et al.\ 2022           &   Jul 2022    &   DIA-NN              &   mTRAQ-3         &   R       &   155     &   \citep{Derks2022-yn}    \\
    \hline \\[-1.8ex]
  \end{tabular}
\end{sidewaystable}


Several conclusions can be drawn from
Figure~\ref{fig:workflows_overview}.  First, one publication
corresponds to one workflow. This variability cannot be explained
solely by different experimental protocols. The computational
pipelines by Schoof et al.\ and Specht et al.\ differ substantially,
while their TMT-based acquisition protocols are closely related
\citep{Specht2021-jm, Schoof2021-pv}, and the computational pipeline
by Liang et al.\ for processing LFQ data \citep{Liang2021-cr} is more
similar to the TMT processing workflow of Williams et al.\ than its
LFQ alternative. Moreover, some publications provide a minimalistic
computational workflow, with only 3 steps, while others perform
extensive processing, with 20 steps. These observations highlight the
lack of consensus and the need to identify critical steps in
computational pipelines. Second, some processing steps are applied at
the peptide level or at the protein level. For instance, Budnik et
al.\ perform normalization at the peptide level, whereas Dou et
al.\ perform normalization at protein level
(Figure~\ref{fig:workflows_overview}I). A clear pattern is that most
pipelines process the data at the protein level, which is questionable
since processing data at an earlier stage could avoid the propagation
of technical artefacts to the protein data \citep{Lazar2016-zl,
  Sticker2020-rl}. Third, a great majority of the methods are taken
from bulk proteomics. We foresee that developing new methods that
account for the properties inherent to single-cell data would
significantly improve the workflows. For instance, batch correction
could benefit from dedicated single-cell methods as the strong
dependency between batch effects and missing data requires robust and
tailored models \citep{Vanderaa2021-ue}. Horizontal integration of
samples from different batches is an active field of research in
single-cell omics \citep{Argelaguet2021-mt} that will probably be
beneficial to the SCP community.

\begin{figure}[!ht]
    \begin{center}
      \includegraphics[width=0.9\linewidth]{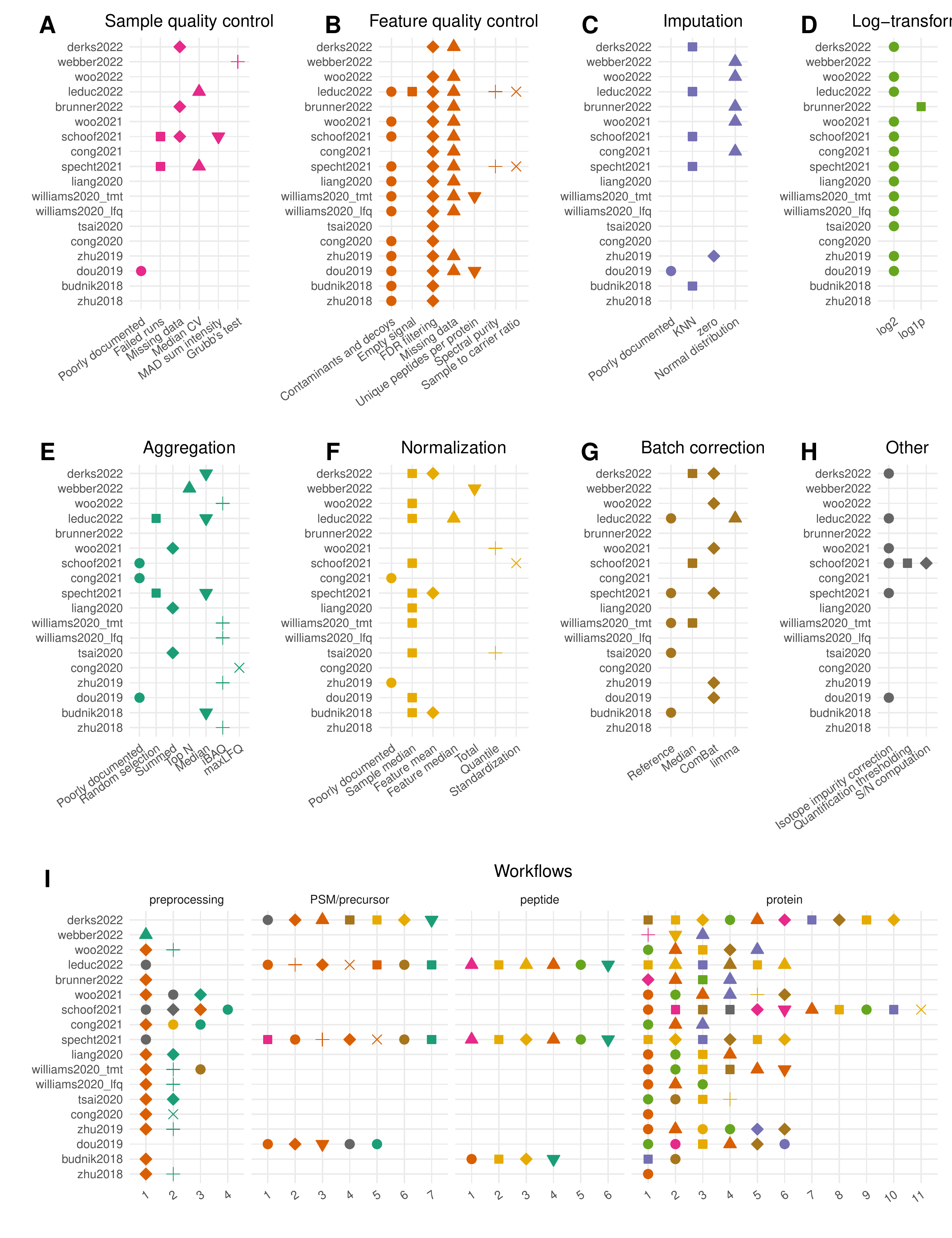}
    \end{center}
    \caption{
      (Continued on the following page.)\label{fig:workflows_overview}
    }
\end{figure}
\begin{figure}[!ht]
  \contcaption{ 
    \textbf{Overview of quantitative data processing workflows. A-H}
    The workflows are split into 8 categories represented by different
    colours.  Each category contains a set of methods that are
    represented by different shapes (see
    Table~\ref{table:method_description} for a description of the
    methods).  Each point indicates which method (column) is
    implemented in which publication (row). Some methods are used in
    several workflows (points align vertically) and some workflows
    used several methods (points align horizontally). \textbf{I}
    Summary of the sequence of processing steps for each workflow.
    Depending on the workflow, the processing steps are implicitly
    executed by the MS data software and applied at the PSM/precursor,
    peptide or protein level. Colours and shapes follow the structure
    from the previous panels. Each horizontal line represents a
    workflow and should be read from left to right.\\
    \rule[0ex]{\linewidth}{0.5pt} }
\end{figure}


Figure~\ref{fig:workflows_overview} highlights the need for a better
understanding on how to process and model SCP data. Identifying which
workflows perform best or demonstrating whether a new workflow
improves performance requires thorough computational benchmarking. The
scRNA-Seq field already offers tools for method benchmarking that
could readily be used for SCP applications \citep{Tian2019-cm,
  Germain2020-ta}. In order to run these tools, the computational
workflows should be accessible to the benchmarking software. Another
key consideration is that benchmarking datasets are required to enable
an objective comparison between computational pipelines. While many
SCP datasets are available from public sources, these data are
provided as different formats. Proper benchmarking necessitates a
standardization of the computational pipelines and the data. In the
next section, we cover the recent developments that attempt to
harmonize quantitative data processing for SCP.

\section{Current solutions for quantitative data processing}


We recently published an R/Bioconductor package called \texttt{scp}
\citep{Vanderaa2021-ue}
(Table~\ref{internet_resources}).  First, \texttt{scp} is thoroughly
documented as we want to facilitate its re-use.  Second, it is
designed as a modular tool where each processing step, such as those
defined in Figure~\ref{fig:workflows_overview} can easily be chained,
and returns a consistent and standardized output format. Third, the
software is part of the Bioconductor project \citep{Huber2015-hj} that
is well known for exemplary coding practices and promotes long term
maintenance. Fourth, \texttt{scp} can be integrated with other tools
that rely on \texttt{QFeatures} and \texttt{SingleCellExperiment}, two
data structures widely used for proteomics and single-cell data
analysis, respectively \citep{QFeatures,Amezquita2020-bf}. Finally,
\texttt{scp} is maintained and improved to include the current
state-of-the-art methods. For instance, it reimplements functionality
from the SCoPE2 script released by Specht et al.
\citep{Specht2021-jm}. Next to \texttt{scp}, Schoof et al.\ developed
\texttt{sceptre}, a python module that implements their computational
workflow \citep{Schoof2021-pv} (Table~\ref{internet_resources}). The
code is well documented, modular and relies on \texttt{scanpy}
\citep{Wolf2018-xo}, a python data structure equivalent to
\texttt{SingleCellExperiment}.  The tool however lacks flexibility as
it was developed primarily to offer a reproducible data analysis
environment. Minor code refactoring could overcome this lack of
flexibility.


Computational solutions require data in order to develop, test and
benchmark individual methods and complete workflows. We therefore also
recently developed another R/Bioconductor package, \texttt{scpdata},
that distributes curated SCP datasets ready for analysis
\citep{Vanderaa2021-ue} (Table~\ref{internet_resources}).
The datasets were retrieved from published work and are accessible
using a single command. The standardization effort provides an
thoroughly annotated and consistent data structure, facilitating data
analysis with tools such as \texttt{scp}. Furthermore,
\texttt{scpdata} relies on Bioconductor's storage services,
\texttt{ExperimentHub} \citep{ExperimentHub}, that offers cloud-based
data access. Easy access and consistent formats enable method
development on a variety of different datasets, avoiding
dataset-specific over-fitting. \texttt{scpdata} can also be used for
benchmarking, although ground truths are missing to perform accuracy
validation.


Standardized data processing tools allow going beyond the reproduction
of existing SCP data analyses, it enables their replication. While
reproduction allows others to regenerate the same results using the
same software or computational setup, replication uses different
software or analysis methods to generate the same, or equivalent
results. Replication, therefore, consolidates our trust in previous
work. Although a replicable analysis does not imply the results are
correct, it guarantees the results do not rely on undocumented steps
or on software peculiarities. For instance, we have shown that the
SCoPE2 analysis script by Specht et al.\ could be fully replicated
using \texttt{scp} and \texttt{scpdata}
\citep{Specht2021-jm,Vanderaa2021-ue}. Replication efforts further
have beneficial side effects. Replication can highlight hurdles that
prevent accurate data analysis. Continuing with the SCoPE2 example,
our replication study identified batch effects and missing data, and
their dependence, as prominent challenges that future SCP
computational tools will need to tackle. Another beneficial side
effect is that replication studies are easily repurposed for
demonstration. As part of this overview, we offer a website,
\textit{SCP.replication}, with replication studies that demonstrate
the analysis of SCP data using the \texttt{scp} and \texttt{scpdata}
packages (Table~\ref{internet_resources}). It contains several
replication articles, spanning TMT and LFQ protocols, and DDA and DIA
data. We also converted the replication material into openly available
workshop material. The workshop can be run without prior installation
requirements thanks to the Orchestra platform supported by the
Bioconductor project.

\section{Different workflows lead to different results}

\begin{figure}[!ht]
    \begin{center}
      \includegraphics[width=0.98\linewidth]{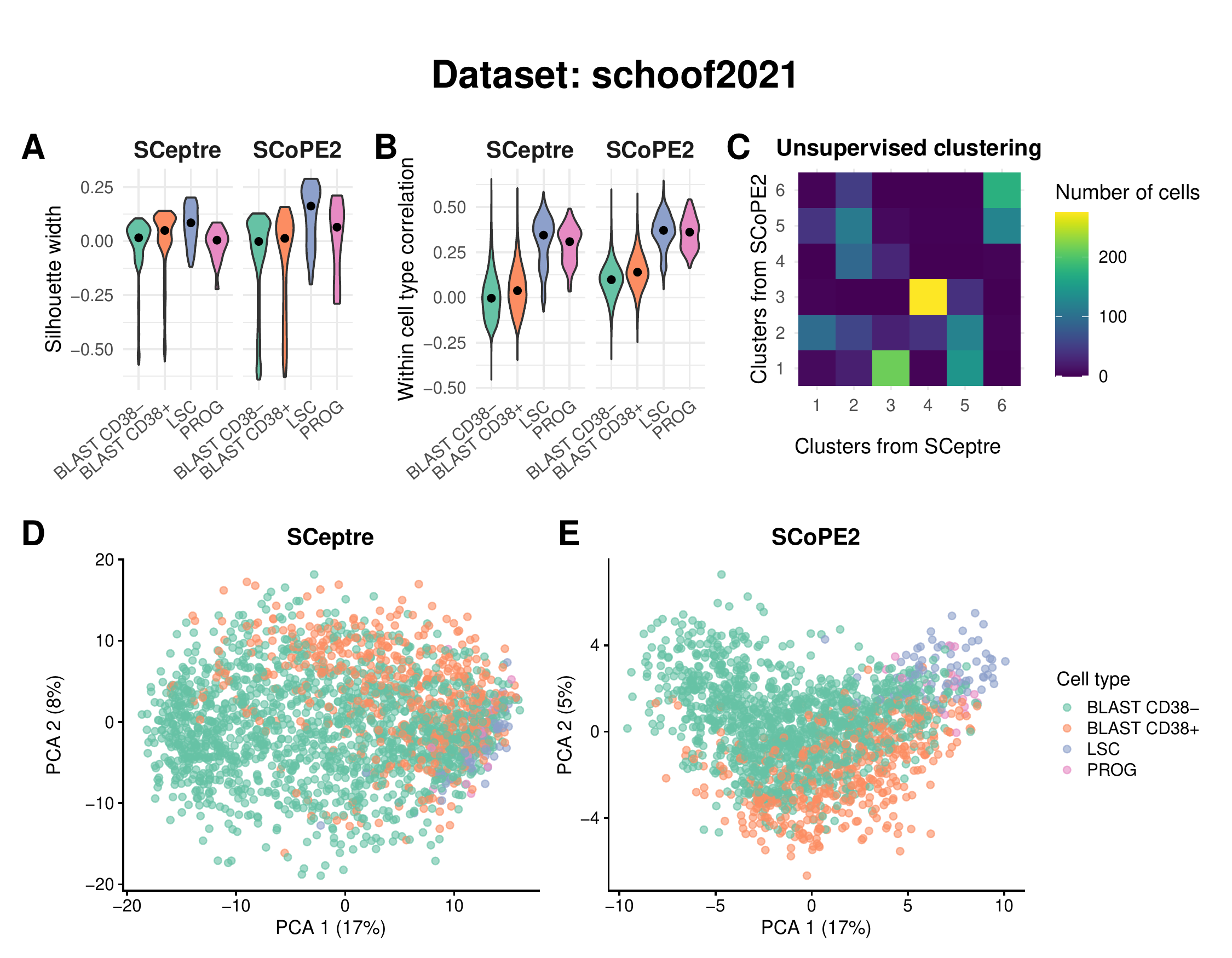}
    \end{center}
    \caption{\textbf{Impact of quantitative data processing workflows
        on the Schoof et al.\ 2021 dataset}. All results presented in
        this figure were computed from the protein data processed by
        the corresponding workflow.  \textbf{A} The silhouette widths
        provide a measure of cell type consistency. Cell types are
        defined based on the known labels provided with the data. The
        silhouettes were computed using the Jaccard similarity on the
        shared nearest neighbour graph (K = 15).  \textbf{B} Pearson
        correlations are computed between all cells with the same cell
        type label and provide a measure of protein quantification
        consistency.  \textbf{C} Unsupervised clustering is performed
        using Louvain clustering \citep{Blondel2008-cx} on a shared
        nearest neighbour graph (K = 15). The heatmap illustrates the
        cell distributions across the clustering results computed from
        the SCoPE2 and the SCeptre workflow.  Frequency is given as
        the number of cells.  \textbf{D,E} The first 2 principal
        components of the protein data.  Colours indicate the cell
        type labels.\label{fig:comparison_schoof2021} }
\end{figure}
\begin{figure}[!ht]
    \begin{center}
      \includegraphics[width=0.98\linewidth]{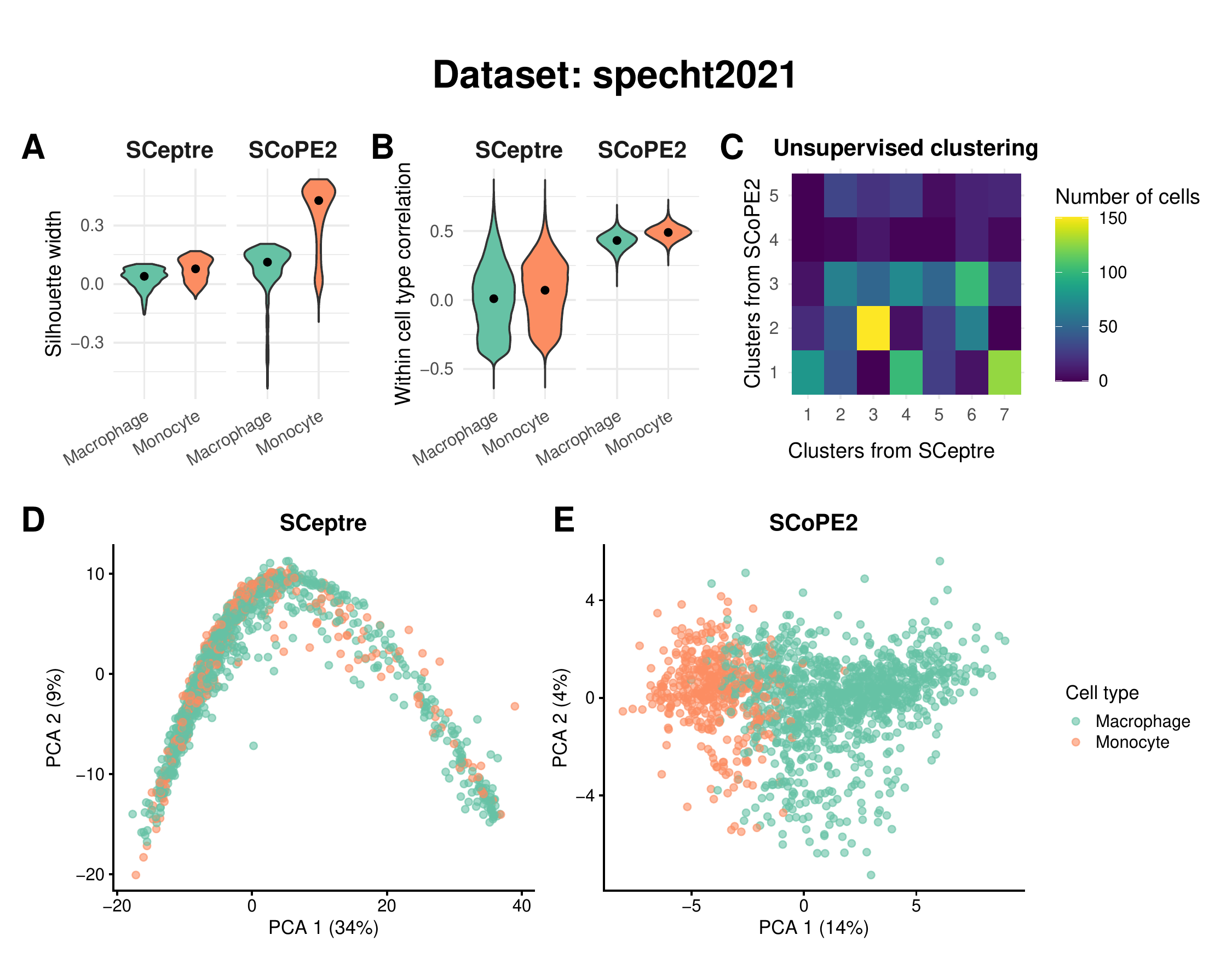}
    \end{center}
    \caption{ \textbf{Impact of quantitative data processing workflows
      on the Specht et al.\ 2021 dataset.} All results presented in
      this figure were computed from the protein data processed by the
      corresponding workflow.  \textbf{A} The silhouette widths
      provide a measure of cell type consistency. Cell types are
      defined based on the known labels provided with the data. The
      silhouettes were computed using the Jaccard similarity on the
      shared nearest neighbour graph (K = 15).  \textbf{B} Pearson
      correlations are computed between all cells with the same cell
      type label and provide a measure of protein quantification
      consistency.  \textbf{C} Unsupervised clustering is performed
      using Louvain clustering \citep{Blondel2008-cx} on a shared
      nearest neighbour graph (K = 15). The heatmap illustrates the
      cell distributions across the clustering results computed from
      the SCoPE2 and the SCeptre workflow.  Frequency is given as the
      number of cells.  \textbf{D,E} The first 2 principal components
      of the protein data.  Colours indicate the cell type
      labels.\label{fig:comparison_specht2021} }
\end{figure}

To illustrate the impact of data processing on the analysis outcome,
we compare two computational workflows: SCoPE2, released by Specht et
al., and SCeptre\footnote{We use \texttt{sceptre} (all lowercase) to
refer to the python library, and we use SCeptre (uppercase ``SC'') to
refer to the computational workflow.}, released by Schoof et al.\ We
retrieved the data from \texttt{scpdata} and recreated both pipelines
with \texttt{scp}. SCeptre uses a custom implementation for batch
correction that is provided by the \texttt{sceptre} library. Thanks to
two R packages, \texttt{reticulate} \citep{reticulate} and
\texttt{zellkonverter} \citep{zellkonverter}, we could easily
integrate the python utilities to \texttt{scp}. We then ran the two
workflows on the two datasets and compared the results in
Figure~\ref{fig:comparison_schoof2021} and
Figure~\ref{fig:comparison_specht2021}.  Both the cell type
consistency, given by the silhouette widths
(Figure~\ref{fig:comparison_schoof2021}A and
Figure~\ref{fig:comparison_specht2021}A), and the within cell type
correlation distributions (Figure~\ref{fig:comparison_schoof2021}B and
Figure~\ref{fig:comparison_specht2021}B) are affected by the
computational workflow. The effect is most visible on the principal
component analysis (PCA) plots for the Specht et al.\ 2021 dataset
(Figure~\ref{fig:comparison_specht2021}D, E). The data
processed with the SCeptre workflow are organized in a horseshoe shape
in lower dimensions. This effect is commonly attributed to the
presence of a latent continuous variable or gradient, and a careful
data exploration revealed the presence of residual batch effects
(Figure~\ref{fig:batch_effect}D). Unsupervised clustering of the
processed protein data leads to different groups, even though we used
identical methods and parameters
(Figure~\ref{fig:comparison_schoof2021}C and
Figure~\ref{fig:comparison_specht2021}C).  The number of identified
clusters differs between the data processed by SCoPE2 and SCeptre.
Furthermore, some clusters from one workflow are scattered throughout
clusters from the other workflow.  Unsupervised clustering is used to
identify groups of cells from which to infer a functional state. So,
different clustering results can lead to different biological
interpretation. To objectively quantify the performance improvement
between the two workflows, we need controlled designs with known
expectations. Benchmarking efforts using mixture designs have already
been performed for scRNA-Seq \citep{Tian2019-cm}. Tian et al.\
acquired both intact single cells and diluted bulk lysates from 3 cell
lines mixed at different proportions and different quantities. These
data were used to assess the ability of computational pipelines to
retrieve the original design.  However, increased performance of a
computational workflow on a single data set is not sufficient.
Different types of SCP data exist (LFQ and TMT data, DDA and DIA data,
orbitrap and time of flight data,\ldots) and computational workflows
may not generalize well for all SCP protocols and data.
In other words, we do not expect a single
computational workflow to perform optimally for all SCP datasets as
different workflows may focus on different characteristics in the
data, such as the pattern of missing values. Therefore, strengths
and weaknesses of computational pipelines need to be identified and
documented. To evaluate this, a community effort could replicate the
mixture design using the different SCP protocols as already seen for
scRNA-Seq \citep{Mereu2020-vk}. This approach would allow to further
assess robustness of the workflows on different types of SCP data.
Although published SCP datasets are available and, as demonstrated
above, they can be processed using standardized software, we are still
lacking the data needed to quantify the performance of computational
workflows.

\section{Lessons learned from replication of SCP data analysis}

When building the replication studies, we faced several practical
challenges regarding the computational analysis of SCP data. In this
section, we provide several recommendations that we hope will help
practitioners to improve and facilitate future SCP data
analyses. While several of these lessons are applicable to bulk
proteomics, we focus on SCP examples here.

\subsection{Complex analyses require suitable tools}


Several search engines for raw MS data identification have been
applied to SCP data: MaxQuant/Andromeda, SEQUEST, MS-GF+, MSFragger
(Table~\ref{experiment_table}).  While MaxQuant/Andromeda is by far
the most popular among those search engines, several authors observed
that it was the worst performing tool in the context of SCP
\citep{Van_Der_Watt2021-mf, Liang2021-cr, Tsai2021-so, Cong2021-qa}.
These observations indicate that one should compare the results of
several search engines in order to maximize the number of reliable
spectrum identifications. Moreover, all search engines applied to SCP
data so far have been developed for bulk proteomics data. Boekweg et
al.\ showed that SCP data have different spectral properties compared
to bulk proteomics data and, hence, the field would benefit from new
search engines developed specifically for SCP data
\citep{Boekweg2021-vl}. Quantitative data processing is also carried
out using different analysis software. Spreadsheet-based and graphical
user-drive software are currently predominant. However, method
development is facilitated by programming languages such as R and
python.  Utilizing programming languages involves a steep learning
curve and is often limited to expert data analysts, but it offers
access to more advanced methods and is a direct and proven solution
for assessable, replicable and re-usable computational analyses.
Finally, an important criterion when choosing software is its
maintenance activity. It has recently been shown that software
accuracy best correlates with the author's commitment to its
maintenance \citep{Gardner2022-is}.

\subsection{Consistent input formats facilitate data analysis}


Formatting input data is a time-consuming and error-prone task when
performing data analysis. To limit this hurdle, \texttt{scp} and
\texttt{sceptre} implement functionality to read structured data
tables. \texttt{sceptre} is designed to read Proteome Discoverer
tables and requires plate annotations and FACS data. \texttt{scp} is
more generic and has been used to read tables from MaxQuant, Proteome
Discoverer, DIA-NN and requires a sample annotation table. Both
implementations require consistent inputs, as provided by software
that export consistent output tables.  Conversely, sample annotation
tables depend on the experimenter. When building the \texttt{scpdata}
packages, we realized that annotation tables are often lacking and
hence needed to be created from the methods section or from the file
names. In other cases, annotations were available through different
files and required heavy data wrangling. This process is
labor-intensive and error-prone. We suggest creating consistent
annotation tables where each row represents a sample (single-cell, TMT
carrier, negative control,~\ldots) and columns represent technical or
biological variables \citep{Gatto2022-kk}. These variables are then
used during statistical modelling to distinguish biological and
technical variability. The annotation tables also require thorough
documentation of the information each column contains. Consistent
input formats streamline data analysis, facilitate the evaluation of
the experimental design (Figure~\ref{fig:batch_effect}B and C), and
provide the information needed for principled data modelling.

\subsection{Beware of confounding effects}

\begin{figure}[!ht]
    \begin{center}
    \includegraphics[width=\linewidth]{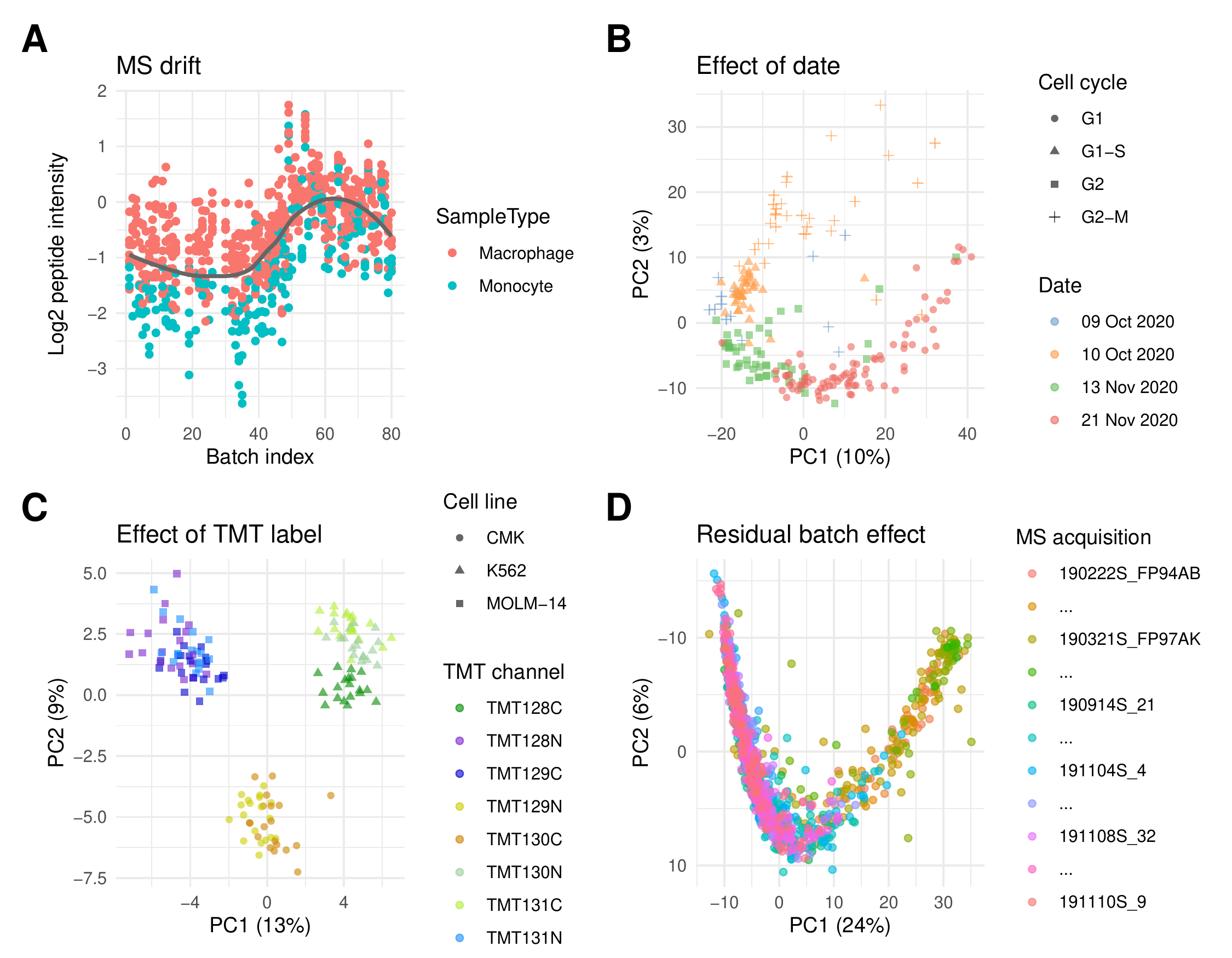}
    \end{center}
    \caption{ \textbf{Confounding effects cause undesired variability.
        A} MS drift for the KLLEGEESR peptide in the Specht et al.\
        2021 dataset \citep{Specht2021-jm}.  The peptide data was
        processed using \texttt{scp} according to the script provided
        along the original paper up to log-transformation
        (Figure~\ref{fig:workflows_overview}I). The batch index is
        ordered by time of acquisition. The MS drift is highlighted by
        a grey line computed using LOESS\@. \textbf{B} PCA plot that
        replicates the results by Brunner et al.
        \citep{Brunner2022-rd}. The protein data was processed using
        \texttt{scp} as an R alternative to the python script provided
        along the original paper
        (Figure~\ref{fig:workflows_overview}I). Shapes represent the
        cell cycle stage; colours represent the day of acquisition.
        \textbf{C} PCA plot that replicates the results by Williams et
        al.\  \citep{Williams2020-ty}. The protein data was processed
        using \texttt{scp} according to the workflow described in the
        Methods section of the original paper
        (Figure~\ref{fig:workflows_overview}I). Shapes represent the
        cell type; colours represent the TMT label.  \textbf{D} Same
        as Figure~\ref{fig:comparison_specht2021}D, but coloured
        according to the MS acquisition batch (n = 149
        batches).}\label{fig:batch_effect}
\end{figure}

All experiments are prone to technical variability and noise, but a
good design of experiment and principled data analysis can disentangle
this undesired technical variability from the desired biological
variability. This is also the case for bulk proteomics, but the
technical challenges are exacerbated when dealing with single-cell
data \citep{Vanderaa2021-ue}.  Real-life SCP experiments require the
acquisition of over hundreds or thousands of cells spread across many
MS runs. Each acquisition run is prone to technical factors that
influence the quantification results. For instance, the MS signal
drift arises from a continuous distortion of the signal between
sequential runs, as already described for bulk proteomics
\citep{Cuklina2021-kf}. Figure~\ref{fig:batch_effect}A confirms that
MS drift is also present in SCP data. Differences between cells over
acquisition time are higher than the differences between cell types at
each time point. A careful design of experiment has spread the two
cell types over time and hence the biological effects can be decoupled
from MS drift thanks to batch correction or statistical modelling.
Neglecting this technical effect can have dramatic consequences. As an
example, Figure~\ref{fig:batch_effect}B depicts an SCP experiment
where single cells are blocked at one of 4 different cell division
stages. Unfortunately, these 4 categories were acquired sequentially,
confounding desired biological and unwanted technical sources of
variation, and impairing deconvolution of the technical and the
biological variability.  When conducting multiplexed experiments, one
must keep in mind that the labels also influence single-cell
quantification \citep{Schoof2021-pv}. This effect was overlooked in
Figure~\ref{fig:batch_effect}C where each TMT tag is assigned to only
one cell type. Again, this impairs computational modelling of the TMT
effects although in this case the biological variance is more
important than the variance associated with the TMT label.

To overcome such confounding effect, it is crucial to carefully design
an experiment using an adequate statistical blocking scheme and
collect data about any technical factors that may influence the
results of an experiment (such as LC-MS/MS maintenance, the type of
instrument used, the multiplexing labels, the sample preparation
batch, the lab that performed the experiment or cell culture batch),
that can interact with known biological factors should also be
gathered, such as cell line, subject ID or tr  eatment condition.
Single cells should then be randomized across all the identified
factor levels. \citep{Schoof2021-pv, Specht2021-jm,Petelski2021-ah,
Gatto2022-kk}. Unfortunately, technical constraints may not allow for
randomized designs. For instance, precious samples from patients may
need to be processed on-the-fly. Hence, the patient identity and their
clinical phenotype will inevitably be correlated with other technical
factors. In scRNA-Seq, pseudo-bulking has been successfully applied to
perform differential expression analysis when the experimental
condition is correlated with the subject. Pseudo-bulking consist in
aggregating cells belonging to the same individual after
identification and separation of the cell (sub-)populations
\citep{Lun2017-ir, Crowell2020-mn}. However, how to aggregate
proteomics data is still to be explored. Another alternative is to use
linear mixed models. Although these models are computationally more
expensive, they overcome the need for aggregation by modelling protein
estimates from the peptide data \citep{Goeminne2020-op}. Finally,
dedicated efforts are required to better monitor and control batch
effects in SCP data.

Finally, it is important to validate the batch correction and exclude
residual batch effects. For instance, we observed that the cell types
in the Specht et al.\ 2021 dataset cannot be separated in lower
dimensions when processed by the SCeptre pipeline. This is because most
of the variability is explained by residual batch effects
(Figure~\ref{fig:batch_effect}D) indicating that the batch correction
method implemented in SCeptre is not suited for this dataset. We
recommend exploring the effect of technical variables in lower
dimension to offer an intuition on residual batch effects. Batch
correction assessment can be adapted from bulk proteomics, such as
comparing correlations within batches and within conditions or
correlations between unrelated peptides and peptides from the same
protein. We refer to~\cite{Cuklina2021-kf} for a thorough discussion.

\subsection{Quantitative data processing depends on downstream analyses}

The purpose of quantitative data processing is to prepare the data for
downstream analyses. Downstream analyses process data into
interpretable statistical results that in turn can lead to new
biological knowledge. Several approaches have successfully been
applied to SCP data. Dimension reduction condenses the data in fewer
variables. These data embeddings are often used for data
visualization, clustering or trajectory analysis. Differential
abundance or differential detection tests identify proteins whose
abundance are statistically different given a distribution model
between experimental conditions or cell clusters. For the latter,
significant proteins are called marker proteins that can be used to
perform cell type identification. Trajectory analysis infers
differentiation or response tracks in the data by estimating a
pseudo-timeline.  It was recently speculated that SCP technologies
enable the study of direct protein regulatory interactions, opening
the analysis to the discovery of new regulatory mechanisms using an
untargeted approach \citep{Slavov2022-fo}. Quantitative data
processing and downstream analysis could also be combined in a single
statistical framework. While no such methods have been developed for
SCP data yet, there are several examples from the scRNA-Seq field that
could be adapted to SCP\@. For instance, scVI \citep{Lopez2018-ye} or
ZINB-WaVE \citep{Risso2018-sa} implement a modelling procedure that
performs normalization, batch correction, imputation and dimension
reduction as part of the same fitting process. Furthermore, scVI
offers a Bayesian approach to perform hypothesis testing directly from
the estimated model parameters. While these compelling modelling
procedures get rid of many processing steps, they still require a
thorough sample quality control and feature selection. Whatever the
chosen downstream method is, quantitative data processing workflows
must match the underlying assumptions and data distributions. For
instance, dimension reduction using PCA requires data imputation, but
dimension reduction using non-iterative partial least squares (NIPALS)
offers a similar alternative that is robust against missing values,
hence does not require imputation \citep{Andrecut2009-sm,
  Vanderaa2021-ue}.  Quantitative values need to be batch-corrected
when running a t-test otherwise the results will become biased and
inaccurate. However, when using linear regression, technical factors
can be directly included as part of the model and do not require
previous batch correction \citep{Ritchie2015-io}.

\section{Concluding remarks}


While standardized SCP protocols are applied outside the pioneering
labs \citep{Petelski2021-ah, Leduc2022-cc, Liang2021-cr, Tsai2021-so},
computational workflows to process SCP data still lack any form of
standardization. The overwhelming diversity of pipelines makes it
difficult to make informed decisions as how to analyse SCP data.  We
provide important guidelines to orient the design of the data
analysis. First, SCP designs and data analyses are complex, and
analysis tools should be carefully chosen. Second, robust data
analysis relies on consistent and standardized data
formats. Standardized data structures should facilitate sample
annotation biological and technical factors that influence data
acquisition. Third, accounting for batch effects is essential to avoid
assigning biological discoveries to technical variation, especially
for SCP experiments, comprising ever-increasing numbers of single cell
samples. Finally, the processing of quantitative data highly depends
on the research question at hand and, hence, on the downstream
analysis to perform. It is not possible to define a good computational
workflow without defining the task to accomplish.

More work is required to offer clear answers on how to set up optimal
SCP experimental designs and associated computational pipeline. The
field still lacks understanding of the impact of each processing step
on the final results. Workflows are still built based on empirical and
arbitrary decisions. As the technology gains in momentum and more
groups start to embrace SCP, setting more complex designs,
standardized and benchmarked computational pipelines are needed to
guarantee sound data interpretation. Indeed, strong data analyses
principles and frameworks will enable the technology to reach its full
potential.  Low quality results generated by flawed analysis practices
could penalize the field instead of incentivize for better
analysis. \texttt{scp}/\texttt{scpdata} and \texttt{sceptre} represent
strong foundations that can support computational benchmarking
efforts.



\paragraph{Conflict of Interest Statement}

The authors have no relevant affiliations or financial involvement
with any organization or entity with a financial interest in or
financial conflict with the subject matter or materials discussed in
the manuscript. This includes employment, consultancies, honoraria,
stock ownership or options, expert testimony, grants or patents
received or pending, or royalties.



\paragraph{Acknowledgments}

This work was funded by a research fellowship of the Fonds de la
Recherche Scientifique-FNRS\@.


\paragraph{Data Availability Statement}

All data used to create the figures in this article are available from
the \texttt{scpdata} package \citep{scpdata}. The R code to reproduce
the figures is available at:
\url{https://github.com/UCLouvain-CBIO/2022-scp-data-analysis}.



\bibliographystyle{plainnat}
\bibliography{ref,software}
\clearpage



\section*{Internet resources}

\begin{table}[h]
  \caption{ \textbf{List of SCP data, computational and didactic
    resources.\label{internet_resources}} }
  \footnotesize
  \begin{tabular}{p{0.14\linewidth} p{0.14\linewidth} p{0.08\linewidth} p{0.6\linewidth}}
    \\[-1.8ex]
    \hline\hline
    \\[-1.8ex]
    & Tool & Language & Description and link \\
    \hline \\[-1.8ex]

    \textbf{Data }& \texttt{scpdata} & R &
    R/Bioconductor package providing published SCP datasets ready
    for analysis in R. \\
    & & & https://bioconductor.org/packages/release/data/experiment/html/scpdata.html \\
    \hline \\[-1.8ex]

    \textbf{Processing} & \texttt{scp} & R &
    R/Bioconductor package to process quantitative SCP data,
    supporting all steps and methods depicted in
    Figure~\ref{fig:workflows_overview}.\\
    & & & https://bioconductor.org/packages/release/bioc/html/scp.html \\
    \cline{2-4}

    & \texttt{sceptre} & Python &
    Python package extending the functionalities of Scanpy to analyse
    multiplexed SCP data. It implements the Schoof et al.\ 2021
    workflow in Figure~\ref{fig:workflows_overview}.\\
    & & &
    https://github.com/bfurtwa/SCeptre\\
    \hline \\[-1.8ex]

    \textbf{Reproduction} & SCP.replication & R &
    Website with 8 SCP data analysis articles that replicate published
    studies.\\
    & & &
    https://uclouvain-cbio.github.io/SCP.replication \\
    \cline{2-4}

    & SCeptre \mbox{notebooks} & Python &
    Code repository containing a set of jupyter notebooks that
    reproduce the results presented in Schoof et al.\ 2021.\\
    & & &
    https://github.com/bfurtwa/SCeptre/tree/master/Schoof\_et\_al/code \\
    \cline{2-4}

    & SlavovLab code & R &
    Set of code repositories that reproduce the results published by
    the Slavov lab.\\
    & & &
    https://github.com/Single-cell-proteomics \\
    \cline{2-4}

    & Brunner code & Python &
    Code repository with a jupyter notebook to reproduce the results
    in Brunner et al.\ 2022.\\
    & & &
    https://github.com/theislab/singlecell\_proteomics \\
    \hline \\[-1.8ex]

    \textbf{Tutorials} & SCP workshop & R &
    Online tutorial with hands-on exercises.\\
    & & &
    https://lgatto.github.io/QFeaturesScpWorkshop2021 \\
    \hline\hline \\[-1.8ex]
  \end{tabular}
\end{table}

\clearpage





\setcounter{table}{0}
\renewcommand{\thetable}{S\arabic{table}}%
\setcounter{figure}{0}
\renewcommand{\thefigure}{S\arabic{figure}}%

\section*{Supplementary information}

\subsection*{Supplementary tables}

\begin{longtable}[h]{>{\raggedright}p{0.15\linewidth} >{\raggedright}p{0.155\linewidth} p{0.7\linewidth}}
  \caption{
    \textbf{Description of SCP data processing methods.\label{table:method_description}}
  }
  \small
    \\[-1.8ex]
    \hline\hline
    \\[-1.8ex]
    Step & Method & Description \\
    \\[-1.8ex]
    \hline\hline
    \endhead
    \\[-1.8ex]

    Sample quality control & Failed runs & Remove samples that belong
    to an MS acquisition that failed based on the number of PSMs.\\
    \cline{2-3}
    & Missing data & Remove samples that contain too many missing
    values. Too many is defined by the authors either using the
    percentage missing value per cell or the number of identified
    proteins per cell.\\
    \cline{2-3}
    & Median CV & Remove samples for which the median coefficient (CV)
    of variation is higher than a user provided threshold. CVs are
    computed from peptides belonging to the same proteins and CVs
    per cell across proteins are summarised using the median.\\
    \cline{2-3}
    & MAD sum intensity & Remove samples for which the total protein
    signal intensity per cell falls outside the median absolute
    deviation (MAD). Briefly, the summed intensities are computed for
    each cell. The median of the absolute differences between the
    summed intensities and the computed median is the MAD.\@ All cells
    that have their summed intensity higher than the median plus MAD
    or lower than the median minus MAD are removed. \\
    \cline{2-3}
    & Grubb's test & Iteratively remove outlying cells based on the
    ratio of the absolute deviation to the mean of all cells and the
    standard deviation. Outlying cells are remove based on a
    significance threshold ($\alpha = 0.05$).\\
    \hline

    Feature quality control &  Contaminants and decoys & Remove
    identified peptides that match the contaminant and/or the decoy
    database.\\
    \cline{2-3}
    & Empty signal & Remove features for which no signal is recorded
    across all single cells.  \\
    \cline{2-3}
    & FDR filtering & Remove peptides or proteins that have an
    associated false discovery rate (FDR) greater than a user provided
    threshold, usually 1\%. FDR is computed by the search engine
    during peptide identification. \\
    \cline{2-3}
    & Missing data & Remove peptides or proteins that contain too many
    missing values. Too many is defined by the authors either using
    the percentage missing value per feature or the number of cells
    the feature was identified in.\\
    \cline{2-3}
    & Unique peptides per protein & Remove proteins that were
    identified and quantified from less than 2 unique peptides.\\
    \cline{2-3}
    & Spectral purity & Remove PSMs that have a low spectral purity,
    that is the proportion of signal from the identified peptide
    compared to signal from background or co-eluting peptide. Spectral
    purity is computed by the search engine during peptide
    identification.\\
    \cline{2-3}
    & Sample to carrier ratio & Remove the PSMs for which the ratio
    between the average intensity across single-cells over the signal
    in the carrier channel is higher than a user provided threshold,
    usually 10\%.\\
    \hline

    Imputation &  KNN & Missing values are imputed using the k-nearest
    neighbours (KNN) averaging. In other words, for each cell, the
    missing value for a protein is replaced with the average intensity
    of K most similar cells (in Euclidean space) for which that
    protein is not missing. K is defined y the user or optimized by
    maximizing the silhouette width between known cell types.\\
    \cline{2-3}
    & zero & Missing values are replaced by a zero value. \\
    \cline{2-3}
    & Normal distribution & Missing values are replaced by values
    sampled from a normal distribution. The mean of the distribution
    is defined as the global mean of non-missing values downshifted by
    a user provided value, and the standard deviation of the
    distribution is set by the user as well.\\
    \hline

    Log-transform & log2 & Replace quantitative values by the $log_2$
    of the values.\\
    \cline{2-3}
    & log1p &  Replace quantitative values by the $ln$ of the values
    with a pseudo count of one.\\
    \hline

    Aggregation & Random selection & Pick one feature to represent the
    aggregated feature. This method was used to aggregate PSMs to
    charged peptides. In this context there are about 0.1\% of charged
     peptides that originate from more than one PSM.\@\\
    \cline{2-3}
    & Summed & Combine quantitative data from multiple features by
    summing their intensities.\\
    \cline{2-3}
    & Top N & Combine quantitative data from multiple features by
    summing intensities of the N most intense features.\\
    \cline{2-3}
    & Median & Combine quantitative data from multiple features by
    taking the median of their intensities.\\\\
    \cline{2-3}
    & iBAQ & intensity based absolute quantification (iBAQ) combines
    peptides into protein quantifications by summing the peptide
    intensities divided by the number of theoretically observable
    peptides.\\
    \cline{2-3}
    & maxLFQ & Protein quantifications are computed from peptide
    quantification using delayed normalization and maximal peptide
    ratio extraction. See~\cite{Cox2014-bs} for detailed information
    on the method.\\
    \cline{2-3}

    Normalization & Sample median & Quantitative values for each
    single cell are divided or subtracted by the median for that
    cell.\\
    \cline{2-3}
    & Feature mean & Quantitative values for each feature are divided
    or subtracted by the mean for that feature.\\
    \cline{2-3}
    & Feature median & Quantitative values for each feature are divided
    or subtracted by the median for that feature.\\
    \cline{2-3}
    & Total & Quantitative values for each feature are divided
    or subtracted by the median for that feature.\\
    \cline{2-3}
    & Quantile & Quantitative values are transformed so that the
    qualtile distribution are aligned between all single cells.\\
    \cline{2-3}
    & standardization & Quantitative values for each feature are
    subtracted by the average and divided by the standard deviation.\\
    \hline

    Batch correction & Reference & For all cells within a batch and
    for each feature, the quantitative data are divided or subtracted
    by the reference samples within that run.\\
    \cline{2-3}
    & Median & For all cells within a batch and for each feature, the
    quantitative data are divided or subtracted by the median values
    within that run.\\
    \cline{2-3}
    & ComBat & Batch effects are removed using linear regression with
    an Empirical Bayes framework. The method models the
    quantifications of each protein with respect to a batch factor as
    well as the biological factors to protect from correction.\\
    \cline{2-3}
    & limma & This method is very similar to ComBat. It allows for
    more than one batch factors.\\
    \hline

    Other & Isotope impurity correction & The signal associated to a
    label is corrected by removing the signal associated to impurities
    from the other labels. This is performed using a compensation
    matrix provided by the label manufacturer or from theoretical
    isotypic composition. This is only applicable for multiplexed
    experiments.\\
    \cline{2-3}
    & Quantification thresholding & Any quantitative value below a
    user provided threshold is set to be missing. See~\cite{Cheung2020-wc}
    for more detailed information on how to set the threshold.\\
    \cline{2-3}
    & S/N computation & Convert MS signal intensities to
    signal-to-noise ratio (S/N). The noise signal is estimated by the
    spectrum quantification software.\\
    \hline
\end{longtable}

\clearpage


\end{document}